\def\lesssim{{_ <\atop{^\sim}}}
\def\grtsim{{_ >\atop{^\sim}}}
\def\apj{ApJ}
\def\aj{AJ}

\def\mnras{MNRAS}
\documentstyle[twocolumn,epsfig]{mn}

\newcommand\Mpch{\mbox{$h^{-1}$Mpc}}
\newcommand\hMsun{h^{-1}{\ }{\rm M_{\odot}}}

\newcommand\kpch{h^{-1}{\ }{\rm kpc}}

\newcommand\msunh{\mbox{$h^{-1}$M$_\odot$}}
\newcommand\Msunh{\mbox{$h^{-1}$M$_\odot$}}

\newcommand\kms{\mbox{~{\rm km s}$^{-1}$}}

\newcommand\Vc{\mbox{$V_{\rm m}$}}
\newcommand\etal{et al. }

\newcommand\Colin{Col{\'{\i}}n}

\def\'#1{\ifx#1i{\accent"13\i}\else{\accent"13#1}\fi}

\begin{document}

\title{Density profiles of dark matter haloes: diversity and dependence
on environment}

\author[Avila-Reese et al.]
          {Vladimir Avila-Reese,$^{1,2}$
           Claudio Firmani,$^{1,2}$
			  Anatoly Klypin,$^2$
			  \newauthor
          and
			  Andrey V. Kravtsov$^2$\\
$^1$Instituto de Astronomia, UNAM, A.P. 70-264, 04510 Mexico D.F.\\
$^2$Astronomy Department, New Mexico State University, Box 30001, Dept.
4500, Las Cruces, NM 88003-0001}

   \date{Received ...; accepted ...}

   \maketitle

\begin{abstract}
We study the outer density profiles of dark matter haloes predicted by
a generalized secondary infall model and observed in a dissipationless
cosmological simulation of a low-density flat cold dark matter model
with the cosmological constant. We find substantial systematic variations in
shapes and concentrations of the halo profiles as well as a strong
correlation of the profiles with the environment in which the haloes
are embedded. In the $N$-body simulation, the average outer slope of the
density profiles, $\beta$ ($\rho\propto r^{-\beta}$), of isolated
haloes is $\beta \approx 2.9$, and $68\%$ of these haloes have
values of $\beta$ between 2.5 and 3.8. Haloes in dense environments of
clusters are more concentrated and exhibit a broad distribution of
$\beta$ with an average value higher than the average $\beta$ for
isolated haloes.  For haloes located within half the virial radius of
the cluster from the center values $\beta \approx 4$ are very
common. Contrary to what one may expect, the haloes contained within
groups and galaxy systems are less concentrated and have flatter outer
density profiles than the isolated haloes: the distribution of $\beta $
peaks at $\approx 2.3-2.7$. The slope $\beta $ weakly anticorrelates
with the halo mass $M_h$. The concentration decreases with $M_h$, but
its scatter is roughly equal to the whole variation of this parameter
in the galaxy halo mass range. The mass and circular velocity of the
haloes are strongly correlated $M_h \propto V_m^{\alpha}$ with $\alpha
\approx 3.3$ and $\approx 3.5$ for the isolated haloes and haloes in
clusters, respectively. For $M_h \approx 10^{12}\msunh$ the rms
deviations from these relations are $\Delta $log$M_h=0.12$ and 0.18,
respectively.  Approximately $30\%$ of the haloes are contained within
larger haloes or have massive companions within three virial radii. The
companions are allowed to have masses larger than $\sim 0.3$ the mass
of the current halo. The remaining $70\%$ of the haloes are isolated
objects. We find that the distribution of $\beta$ as well as the
concentration-mass and $M_h-V_m$ relations for the isolated haloes
agree very well with the predictions of our {\it seminumerical\/}
approach which is based on a generalization of the secondary infall
model and on the extended Press-Schechter formalism.

\end{abstract}

\section{Introduction}

The dark matter (DM) haloes are thought to be objects within which
luminous galaxies form and evolve. Thus, the properties and
evolutionary features of the observed galaxies should be related to
their haloes.  According to the hierarchical scenario, the DM haloes form
via collapse of primordial density fluctuations in the
expanding Universe. Cosmological $N$-body simulations provide a direct
way to study this process. Nevertheless, only recently the simulations became
accurate enough to resolve the internal structure of the galaxy-size
haloes produced in these simulations. Navarro, Frenk, \& White (1996,
1997; hereafter NFW) found that for a large
range of masses and background cosmologies the density profiles 
of equilibrium haloes are well
approximated by the {\it universal\/} profile:
 
\begin{equation}
\rho(r) =\frac{\rho _s}{r/{r_s}(1+r/{r_s})^{2}}, 
\end{equation}

\noindent where $r_s$ is a characteristic radius, and $\rho _s$ is a 
characteristic density. NFW found that these two parameters
are connected,
and that the remaining free parameter depends only on mass. This free
parameter is the concentration $c_{\rm NFW}=r_v/r_s$, where
$r_v$ is the halo virial radius.  This conclusion was later extensively
tested and confirmed with other numerical simulations (e.g., Cole \& Lacey
1996; Kravtsov, Klypin \& Khokhlov 1997; Moore \etal  1998; Bullock
\etal  1999). Studies based on analytical and seminumerical methods also have 
shown that eq. (1) is a good approximation for the density profiles of
typical equilibrium cold dark matter haloes (e.g., Syer \& White
1998; Avila-Reese, Firmani, \& Hern\'andez 1998; Salvador-Sol\'e,
Manrique \& Solanes  1998; Raig, Gonz\'alez-Casado \& Salvador-Sol\'e 
1998; Henriksen \& Widrow 1999; Nusser \& Sheth 1998; Kull 1999; Lokas
1999). Nevertheless, the applicability of the NFW profile has some limits.  
Avila-Reese \etal  (1998) (hereafter AFH98)
have shown that the NFW profile describes well only the structure of
typical haloes formed from the most probable hierarchical mass
aggregations histories (MAHs). Depending on the MAH, diverse
density profiles are possible. There is another limit on the
applicability of the NFW profile: it is formally valid only for
isolated haloes, which are in equilibrium and which do not have large
companions. These limits on the applicability of the NFW profile point
out to the necessity to find in the numerical simulations what kind of
profiles have the haloes that deviate from the NFW case and how the 
different environments in which the haloes are embedded influence on
their density profiles.

Recently, Jing (1998,1999) has studied the density profiles of
hundreds of DM haloes in high-resolution $N$-body simulations. He
concluded that although the NFW profile describes the structure of a
large fraction of haloes in equilibrium, there are other haloes whose
density profiles deviate from the NFW shape. In particular, the haloes
with significant internal substructure show large deviations from the
NFW profile. Bullock \etal (1999; see also Jing 1998, 1999) have found
a substantial scatter in the parameter $c_{\rm NFW}$ that may be due
to the fact that the NFW profile does not describe well some of the
profiles. Bullock \etal also find that $c_{\rm NFW}$ depends on the
environment.

 To determine the shape of the most inner parts of the halo density profiles
the simulation should have a very high resolution. Whether this shape is 
$r^{-1}$ as for the NFW profile or not, it is still
matter of debate (Kravtsov \etal  1998; Moore \etal 
1998,1999; Jing 1999). The resolution of the numerical simulations
discussed herein is not sufficient to address the issue of the central
slope.

In this paper we study intermediate and outer regions ($r \grtsim
r_s$) of the halo density profiles of a large sample of DM haloes
obtained in a cosmological simulation of a flat low-density cold dark
matter model with cosmological constant ($\Lambda $CDM).  Details of
the simulation are described in Col\'{\i}n et al. (1999) and Kravtsov
\& Klypin (1999). To identify 
haloes, we use the Bound Density Maxima (BDM) halo finding algorithm
(Klypin
\etal 1999). The main difference with the results presented by Jing
(1998, 1999) is that we study both the isolated haloes and the
satellite galaxy-size haloes residing inside groups and clusters.  As
the result, we are able to examine the environmental distribution of
the galaxy-size DM haloes and the influence of the environment on DM
halo structural properties.

Analytical and seminumerical methods, alternative and complementary to
the expensive cosmological $N$-body simulations, have proved to be
useful. These approaches allow one to follow some particularly chosen
features and phenomena of DM halo formation. Here, the results
obtained from the $N$-body simulations
will be compared and, when possible, will be interpreted in the light
of the predictions of the seminumerical method presented in
AFH98. This method is based on a generalization of the secondary
infall model (see, e.g., Zaroubi \& Hoffman 
1993) and  uses the extended Press-Schechter formalism in order to calculate the 
hierarchical MAHs.

 In $\S$2.1 we briefly describe the numerical simulation and the
algorithm used to identify DM haloes and to obtain their density
profiles. An outline of the seminumerical approach is given in
$\S$2.2. The spatial distribution of the identified haloes according to the
environment is presented in $\S$3. In Section 4 we examine the
outer density profiles of the haloes and the way in which
these profiles change with the
environment. The structural correlations of the haloes as a function of
the environment are analyzed in section 5. The concentration-mass
and mass-velocity relations are presented in $\S$5.1 and
$\S$5.2, respectively. The estimated dispersions for
these relations and the correlation among them are also
presented. In $\S $ 6 we discuss some of the results. The summary and 
conclusions of the paper are given in $\S $7.

\section{Numerical and seminumerical simulations}

\subsection{$N$-body simulations and halo identification algorithm}

Structure of DM haloes only slightly depends on cosmology (e.g., NFW;
Cole \& Lacey 1996; Kravtsov \etal 1997; Avila-Reese 1998; Firmani \&
Avila-Reese 1999a). Therefore, the results for a representative
cosmological model should be sufficient to outline the general
behavior and trends of the structural properties of the DM haloes. In
this study, we use a flat cold dark matter model with the cosmological
constant ($\Lambda$CDM). The model has the following parameters: the
density of matter is $\Omega _0=0.3,$ the density due to the vacuum
energy is $\Omega_\Lambda =0.7,$ the Hubble constant is $H_0=100h\kms
{\rm Mpc}^{-1}$ with $h=0.7$, the amplitude of perturbations on $8\Mpch$
scale is $\sigma _8=1$. The numerical simulation was done using the
ART code described in Kravtsov \etal (1997). Details of the simulation
are presented in \Colin~ \etal (1999). The simulation followed
evolution of $256^3$ particles in a $60\Mpch$ box. The mass of a DM
particle is $1.1\times 10^9\Msunh$. The peak force resolution is
$1.8\kpch$. High mass and force resolution are very important for
survival of DM haloes in dense environments of groups and clusters of
galaxies.

The Bound Density Maxima (BDM) halo identification algorithm (Klypin et
al. 1999) was applied to find the DM haloes. The algorithm locates
maxima of density within spheres of radius $10\kpch$ and then removes
unbound particles. The algorithm produces a catalog of DM haloes
containing coordinates, velocities, and density profile of {\em bound\/}
particles for each halo. The density profile is used to find the
maximum circular velocity $\Vc = (GM/r)^{1/2}$, radius and mass of the
halo.  We define the halo radius $r_h$ as the minimum of the virial
radius $r_v$ and the truncation radius $r_t$. The former is defined as
the radius at which the average density of the system is $\Delta _c(z)$
times the average density of the universe at redshift $z$, where
$\Delta _c(z)$ is determined from the spherical collapse model. For the 
model we use here $\Delta _c(z=0)=334$ (e.g., Bryan \& Norman 1998; 
but see a recent paper by Shapiro, Iliev, \& Raga 1999 where 
a more proper and self-consistent treatment of the spherical 
collapse was carried out; for an Einstein-de Sitter universe they 
obtained that $\Delta _c(z=0)$ is $\approx 11\%$ smaller 
than the standard value). The truncation radius is the 
radius where the spherically
averaged outer density profile flattens or even increases. This radius
marks the transition from the halo to the surrounding environment. Only
a small fraction of the haloes identified in our simulation ($\sim 6\%$)
have $r_t<r_v$. The fraction of truncated haloes is larger for the
non-isolated haloes than for the isolated ones. The mass of the DM haloes
$M_h$ is defined as the mass enclosed within $r_h$.

The BDM algorithm is capable of finding haloes with 20-25 bound
particles. In the simulation there were 9073 identified haloes with
this lower limit on number of particles. Analysis of these haloes
indicates that haloes with maximum circular velocity \Vc~ larger than
90\kms~ (6819) are not affected by numerical effects and/or details of
the halo identification (Gottlober, Klypin, \& Kravtsov
1999). Nevertheless, because we need to find the shape of the density
profile, we restrict ourselves to more massive haloes with
$\Vc>130$\kms. The number of haloes in our final catalog is 3498, which
still is large enough for our purposes. All haloes in the catalog have
more than 200 particles.

\subsection{The seminumerical method}

AFH98 and Avila-Reese (1998) presented an approach to study the
gravitational collapse and virialization of DM haloes formed from the
Gaussian density fluctuations. The first step of the method is to
generate hierarchical MAHs of DM haloes. We use the extended
Press-Schechter approximation based on the conditional probabilities
for a Gaussian random field (Bower 1991; Bond et al. 1991; Lacey \&
Cole 1993). For a given present-day mass, we generate a set of MAHs
using Monte Carlo realizations. We follow the aggregation history of
the main progenitor by identifying the most massive subunit of the
distribution at each time. Then, the gravitational collapse and
virialization of the DM haloes formed with the MAHs is calculated
assuming spherical symmetry and adiabatic invariance during the
collapse with an iterative {\it seminumerical} method. 

This method is based on the secondary infall model (e.g., Zaroubi, \&
Hoffman 1993). This model is modified to allowing non-radial motions
and arbitrary initial conditions (MAHs in our case). The only free
parameter in this approach is the ellipticity of the orbits
$e_0=r_{\rm peri}/r_{\rm apo}$, where $r_{\rm peri}$ and $r_{\rm apo}$
are the pericentric and apocentric radii of an orbit,
respectively. The parameter $e_0$ mainly influences the central
structure of a halo: the more circular are the orbits (larger $e_0$),
the shallower is the inner profile. $N$-body simulations indicate that
$e_0$ is typically 0.1-0.4 in cluster-size haloes (Ghigna \etal
1998). Here, we set $e_0$ equal to 0.15. This is the value for which
the density profile of a halo of $10^{12}M_{\odot}$ produced with an
average MAH has the same profile of an isolated well resolved halo of
the same mass found in our $N$-body simulation.

In order to start the Monte Carlo realizations, we fix the present-day 
halo mass, referred here as the nominal mass $M_{\rm nom}$. At any time,
the outer shells that encompass this mass are still in the 
process of virialization. The mass shells that are already virialized
roughly correspond to those within the virial radius $r_v$ at which the mean
overdensity drops below the critical value $\Delta _c(z)$ given by the 
spherical collapse model. Analyses of haloes identified in numerical
simulations show that at radii smaller than $r_v$
the mater is indeed close to a virial equilibrium (e.g., Cole \& Lacey 1996; 
Eke, Navarro, \& Frenk 1998). At radii between $r_v$ and
2$r_v$ the matter is still falling onto the halo, while at larger radii, 
the matter is expanding with the universe. The mass contained
within $r_v$ is the virial mass $M_v$ which, depending upon the MAH, 
is equal to 0.7-0.9 times $M_{\rm nom}$ (see also Kull 1999). Because in the 
numerical simulations the mass of the haloes is 
defined by $M_v$ (only in a few cases is defined by the mass at the truncation 
radius), for the seminumerical simulation we also use  $M_v$.

\section{The environments of galaxy DM haloes}

We divide haloes into two broad categories. One category is constituted 
by haloes whose centers do not lie within the radius of any other 
halo of equal or larger maximum circular velocity. We shall call these 
haloes {\it distinct}. Note, however, that the distinct haloes may contain 
other smaller haloes within their radii. The other category of haloes is 
residing within radii of haloes of larger maximum circular velocities. The
haloes of this category are further divided into three sub-categories
according to the size of their parent halo. If the parent halo has
maximum circular velocity of $\Vc>600\kms$, $350\kms <\Vc\leq
600\kms$, or $\Vc\leq 350\kms$, we will refer to them as {\it haloes in
clusters}, {\it in groups}, and {\it in galaxies}, respectively.  The
limits which define the circular velocities of the cluster, group, and
galaxy haloes are arbitrary. Nevertheless, they reflect velocity ranges
of real clusters, groups, and galaxies. It should be taken into
account that the maximum circular velocity of galaxy-size systems
typically increases by a factor of 1.2-1.4 due to dissipation in the
baryonic component (AFH98; Mo, Mao \& White 1998).

Distinct haloes may or may not have massive neighbors.  We shall call 
{\it isolated} those haloes that do not have a large companion with
$V_{m}^{\rm comp}>f_V\Vc$ within $3r_h$, where
\Vc~ and $r_h$ are the maximum circular velocity and radius of the
current halo, and we have fixed factor $f_V=0.7$. In $\S 5.2$ we find
that halo mass is related to the circular velocity approximately as $M_h\propto
\Vc^{3.3}$. The constraint on circular velocity of the companion 
corresponds roughly to $M_{h}^{\rm comp} > 0.3M_h$, where $M_h$ is the
total mass of the current halo. Thus, an isolated halo is an object not
contained within other halo and without massive companions up to a
relatively large distance. If a halo is not contained inside other halo 
(distinct) but has at least one massive companion 
($V_{ m}^{\rm comp}>0.7\Vc$) within $3r_h$, we consider it
as belonging to a multiple system. In fact, most of the multiple 
systems ($\sim 80\%$) are just pairs. That is why we shall refer 
to this class as the {\it haloes in pairs}.

Table 1 gives the numbers and percentages of galaxy satellite haloes,
haloes in clusters, and haloes in groups. Only $12.5\%$ of all haloes
belong to the category of haloes {\it contained} inside larger
haloes. This fraction remains almost constant if we include in our
catalog smaller haloes with $\Vc <130\kms$ and with less than 200
particles. With the aim to find the fraction of haloes not
contained inside larger haloes (distinct) but with massive companions 
(multiple or pair systems), we analyze the surroundings of each 
of the distinct haloes in search for companions. We may ask ourselves 
what is the distance
$d_{\rm comp}^{\min}$ to the nearest companion with $V_{m}^{\rm
comp}>0.7\Vc$, where \Vc~ is the maximum circular velocity of the
current halo. In Figure 1 we present differential and cumulative
distributions of $d_{\rm comp}^{\min}$ normalized to the radius $r_h$
of the halo. Although we usually consider only haloes with $\Vc
>130\kms$, companions were allowed to have smaller circular velocities
($>90$\kms). This was done to allow even a small halo ($\Vc \approx
130\kms$) to have a chance to have companions as small as 0.7 of
their own circular velocity. We find that only a small fraction ($\sim
2\%)$ of haloes of the category in study contains a halo of mass larger
than $\sim $0.3 of their mass ($V_{m}^{\rm comp}>0.7\Vc$) within their
total radius $r_h$. Most of the haloes of this category have companions
with $V_{m}^{\rm comp}>0.7\Vc$ as far as 2-4 times their radius. The
isolated haloes, as defined above, constitute the $80\%$ of the distinct 
haloes and the $70\%$ of all the haloes (see Table 1).

The large fraction of isolated haloes found in the numerical simulation 
actually strongly depends upon the parameter $f_V$. In Figure 2 we plot
the fraction of isolated haloes with respect to all the haloes as a
function of $f_V$. Because we prefer to limit our catalog only to
haloes with \Vc $>$ 90\kms~ (this is the minimum velocity allowed for
the companion haloes), the limit on \Vc~ of the isolated haloes has
to be increased when $f_V$ decreases in order for the sample to remain
complete. That is why as $f_V$ decreases we should use catalogs
with larger limits on $V_m$. The number of isolated haloes significantly
decreases when the minimum mass of the companions decreases. Our
results agree with the halo-halo correlation function for isolated
haloes.  

The fractions of objects in different systems found at $z=0$ in our
numerical simulation roughly agree with what is observed in the
Universe: $60\%-70\%$ of galaxies are in the field (most of them are
disc galaxies), $30\%-40\%$ are in groups (e.g., Ramella, Geller,
Huchra 1989; Nolthenius, Klypin, \& Primack 1994), and $5\%-10\%$ are
in clusters (Bahcall 1988). It should be noted that some of the pair
haloes ($\sim 17\%$ of all haloes in the sample) might be classified as
small groups composed of two relative large galaxies and a few small
satellites. Moreover, as it was mentioned above, $\sim 20\%$ of these
haloes actually have more than one massive companion, i.e. they form
multiple systems.

\section{Density profiles}

\subsection{$N$-body simulations}

The mass resolution in our simulation ($m=1.1\times 10^9\Msunh$) is not
sufficient to resolve central parts of most of our haloes. As the
result, we focus on the structure of the outer profile. Our first
question is whether the halo density profiles have the shape of the NFW
profile.  AFH98 (see also Jing 1998,1999) find that the DM haloes
actually have a {\it range} of density profiles for a given mass where
the average profile may be described by the NFW profile. Using the
large sample of haloes identified in our numerical simulation, we fit
the spherically averaged density profile of the haloes by the following
function:

\begin{equation}
\rho (r)=\frac{\rho _s}{\frac r{r_s}(1+\frac r{r_s})^{\beta -1}}. 
\end{equation}

\noindent This is a generalized NFW profile where the slope $\beta$ of
the outer part of the profile
($\rho (r)\propto r^{-\beta }$ for $r>> r_s$) may be different from the
slope $\beta =3$ of the NFW profile.

The density in the inner regions with less than 50 particles has a
rather large shot noise. Therefore, we use only those bins which have
more than 50 particles inside them. For most of the DM haloes, the
radius from which the halo has more than 50 particles is $\sim 0.3-0.8$
of the radius where the maximum circular velocity $r_m$ is
reached (for the NFW profile the radius $r_m$ is about
$2.2r_s$). Since our interest is in galaxy-size haloes, we additionally
restrict the sample to haloes with $\Vc<350$\kms. This reduces the
number of haloes to 3347 (out of 3498).

The frequency distribution of the parameter $\beta $ obtained for the
galaxy-size haloes is plotted in Figure 3. We have found that $\beta $
does not depend on mass; there is actually an indication for a weak
anticorrelation. The arrows in the horizontal axis, from left to
right, indicate the $16\%$, $50\%$ and $84\%$ of the cumulative
distribution, respectively. In other words, roughly $68\%$ of the DM
haloes have values of $\beta $ between 2.50 and 3.88, where the median
corresponds to $\beta \approx 2.94.$ This result does not depend
strongly on the quality of the fit. In Figure 3 the frequency
distribution of $\beta $ for those profiles that were fitted with an
accuracy better than $(\chi ^2/N_{\rm bins})^{1/2}<7\%$\footnote{ In
our case, the quantity $\chi ^2$ used for the minimization of the
fitting is relative (dimensionless) because we fit the logarithm of
the density ( $\chi ^2=\sum_i^N(\log (\rho _i/
\rho _{an}))^2,$ where $\rho _{an}$ and $\rho _i$ are the analytical
and measured values of the density, respectively} is plotted with the
thin solid curve. Approximately $33\%$ of all the haloes used in this
analysis have a fitting with this accuracy.  The number of bins $N_{\rm
bins}$ varies from halo to halo depending on its size. This is why we
divide $\chi ^2$ by $N_{\rm bins}$ in order to have an estimator of the
goodness of the fit. The distribution becomes only a little narrower
and slightly shifts to smaller values of $\beta $ than in the case when
all the profiles are considered.

An error analysis of the slope $\beta$ is important in our case because
the number of particles in an average halo is not very large. We
roughly estimate limits of the error in the following way. For a given
halo, we generate a set of density profiles drawn from an ensemble of
profiles with the mean of the original halo profile and with deviations
defined by the Poisson noise due to finite number of particles. The set
is used to estimate the errors in $\beta$ produced by finite number of
particles in the halo. Specifically, for each radial bin of a halo
profile we find the number of particles in the original halo and then
perturb it assuming the Poissonian distribution of the particles in the
bin.  Repeating this procedure for every bin and several times for each
halo, we get a set of density profiles for a given halo. Applying the
fitting procedure to each of these density profiles, we obtain a set of
values for the fitting parameters ($\beta ,$ $r_s,$ and $\rho _s)$ for
which we can estimate the standard deviations. This method provides a
way to estimate the uncertainty  on $\beta$. 

Having in mind that $\beta $ does not depend on the mass, we have
applied the experiment to three groups of haloes with the slope $\beta
$ around 2.50, 2.94 and 3.88. These values of the slope correspond to
the $16\%$, $50\%$ and $84\%$ of the cumulative distribution of this
parameter. For each group we selected dozens of haloes, and for each
halo we applied the Monte Carlo experiment 30 times. The average
standard deviations of the parameter $\beta $ for each group are shown
in Figure 3. Note that the dispersion of $\beta$ increases with $\beta
$. This is expected because for haloes of a fixed mass, the number of
particles in external bins is smaller when $\beta $ is
larger. Therefore, the Poisson noise of external bins is larger for
larger $\beta $. The dispersion of the other two parameters,
particularly $\rho _0,$ have the opposite trend. Thus, a significant
contribution to the uncertainty in $\beta,$ particularly when $\beta $
is large, is probably due to the relatively small number of halo
particles (see also Figure 4a).  A minor contribution to the
uncertainty in $\beta$ may be due to the fitting technique. The Monte
Carlo experiment that we have applied to the haloes, also can be viewed
as a procedure to produce small deformations in the density
profiles. In a few cases the fitting technique can give completely
different values of $\beta$ for a set of these profiles. These cases
typically happens when the density profile abruptly changes from a
very shallow slope to a very steep slope.  In these cases the scale
radius $r_s$ is typically fixed at very large and unphysical values.

\subsection{Dependence of the outer halo density profiles on environment}

The outer part of the density profile eq. (2) is described by the
parameter $\beta .$ We find that this parameter depends on the halo
environment. In Figure 4 the distribution of $\beta $ is shown for
haloes in different environments. The frequency of haloes is defined with
respect to the number of objects in the given category
(environment). Because most of the haloes are isolated, the distribution
for all the haloes presented in Figure 3 remains almost the same for the
isolated haloes. The distribution corresponding to haloes with
$\Vc>350$\kms~ plotted in Figure 4a is slightly narrower and it has
lower amplitude at high values of $\beta .$ But the difference is
small. Thus, it appears that large and galaxy-size haloes have
similar distribution of the slope $\beta$.  The differences are
probably due to the fact that more massive haloes have more particles,
and, thus the scatter on the outer density profiles for them is
smaller, particularly when $\beta $ is large.

The external slope for pair and galaxy satellite haloes in most cases
is $\beta \approx 2.2-2.6$ (Figure 4b), which is shallower than the 
slope in the NFW profile. For
the haloes in groups the distribution of $\beta $ is wider and shifted
to larger values of $\beta $ with a maximum frequency around $\beta
=2.6-2.8$. In the case of galaxy haloes in clusters, the
distribution of $\beta $ is even wider than in the other environments,
with a maximum in $\beta =3.1-3.4.$ If we select only the galaxy haloes
contained within half the total radius of the clusters, then we find that
values of $\beta \approx 4.0-4.4$ are more frequent (dashed line in
Figure 4c). It should be considered, however, that the uncertainty in
the determination of $\beta $ is large when $\beta $ is large (see the
error bar that accounts for the average standard deviation estimated
for the cluster haloes with values of $\beta $ near 3.9). In any case,
the trend of the parameter $\beta $ with the environment is clear. 

With the aim to visually judge the quality of the fitting, in Figure 5
we plot the spherically averaged density profiles and the
corresponding fitting to eq. (2) for isolated haloes, galaxy satellite
haloes, group haloes, and cluster haloes. In this Figure we have plotted
for each category three randomly chosen haloes, each one from a given
range of masses, and with values of $\beta $ around the corresponding
maximum of its distribution. Except for a few cases, the fitted
density profile describes very well the structure of the DM haloes. To
show the quality of the fitting even in the cases where the
uncertainties of the fit are high (when $\beta $ is large), we present
the profiles for the haloes in clusters with $\beta $ around 4.0 instead of
around 3.3. In Figure 5, comparing the profiles of the most massive
haloes (upper curves) with the less massive (lower curves), it can be
appreciated how the number of particles influence on the quality of
the result. The profiles of the less massive haloes (less particles)
are noisier than those of the more massive haloes. Therefore, the
fitting for the former is more uncertain than for the latter.

We also use another way to fit the halo density
profiles in order to check our results. Instead of leaving the
parameter $\beta $ free (see eq. (2)), it was fixed to two different
values $\beta =3 $ and $\beta =4$, where the former corresponds to the
NFW profile and the latter to the Hernquist profile (Hernquist 1991). As
already was mentioned, the quantity $\chi ^2/N _{\rm bins}$ may serve
to some degree as a criterion of the goodness of the fit. In Figure 6
the values of ($\chi ^2/N _{\rm bins})^{1/2} $ obtained for the NFW
and Hernquist profiles are compared for two samples: the isolated haloes
(panel a) and the cluster haloes (panel b). For the former, the NFW
density profile in most of the cases is a better approximation than the
Hernquinst profile. For the latter sample, the density profiles of a
large fraction of haloes are better fitted by the Hernquist profile than
by the NFW. This supports the result that haloes in clusters tend to have  
steeper outer density profiles than the NFW shape.

\subsection{Results from the seminumerical method}

Using the seminumerical approach we produce catalogs of halo profiles for
each chosen mass $M_{\rm nom}$. In order to estimate the slope $\beta$
and the concentration we apply the same fitting procedure used for the 
results of the numerical simulation. Because in the seminumerical approach we
are not able to introduce dynamical effects related to the environment
like the tidal stripping, and due to the assumption of spherical
symmetry, several effects related to non-sphericity, particularly the
major mergers, are not  considered. This is why the haloes
produced in the seminumerical simulations correspond to
the isolated haloes identified in the numerical simulations. In Figure
4(a), the dashed line represents the distribution of the parameter
$\beta$ obtained from the seminumerical approach. As in the numerical
simulations, in this case we also find only a very weak dependence of
$\beta$ with the mass. The Figure 4a shows that both the
simulation and the seminumerical approach produce similar results.

Strictly speaking, the distributions of $\beta $ for the haloes produced in 
the numerical and seminumerical simulations presented in 
the upper panel of Figure 4, do not correspond to the same estimate.
In the case of the numerical simulations, as it was discussed in $\S 4.1$,
the uncertainty due to the relatively small number of particles introduces
an extra scatter on the distribution of $\beta $. The procedure we have
used to calculate this scatter  likely overestimates the errors because it
does not preserve mass of the halo. Moreover, in few cases some
contribution to the calculated scatter can be also due to the ambiguity of 
the fitting technique (see $\S 4.1$).  This also applies for the
haloes in the seminumerical simulation. The intrinsic 
distribution of $\beta $ for the isolated haloes obtained in the numerical 
simulation will be narrower than that presented in the upper panel of 
Figure 4.  Nevertheless, we expect that it does not differ much from the 
distribution obtained in the seminumerical simulations. 

According to the seminumerical approach the differences in the
structure of the haloes are mainly related to the dispersion of the
MAHs. The origin of this 
dispersion is due to the statistical nature of the primordial density
field. Haloes that have larger rates of mass aggregation
at earlier times (early collapse), have density profiles more
concentrated and typically have steeper outer slopes than haloes with
larger mass aggregation rates at later epochs. We find that the outer
slope $\beta $ is particularly sensitive to the behavior of
the MAH at late epochs (close to $z=0$). For example, if the halo
suffers a very late major merger, its outer profile slope $\beta $
will be small ($\beta \lesssim 2.6$). On the contrary, if the mass
aggregation rate is very small for $z \lesssim 1$, then $\beta $ tend 
to be larger than 3.

\section{Concentration and structural correlations of the haloes}

\subsection{Mass vs. concentration}

For the CDM-like power spectra of fluctuations, the hierarchical MAHs
of the DM haloes are such that on average the less massive objects
attain a given fraction of its present-day mass slightly earlier than
the more massive ones. Therefore, on average, less massive haloes are
more concentrated. For the NFW profile, $c_{\rm NFW}=r_v/r_s$ is a
reasonable and physically motivated parameter of
concentration. However, in the case of the more general profile given
by eq.  (2), the scale radius $r_s$ has different physical meanings
for different values of $\beta .$ That is why it is desirable to
define a concentration parameter {\it independent} from the fitting
applied to their density profiles\footnote{As a matter of fact, it is
not possible to characterize the structure of the DM haloes emerged
from a stochastic density fluctuation field with only one (universal)
parameter. The diversity of density or mass profiles associated with
the diversity of MAHs (see e.g., AFH98; Firmani \& Avila-Reese
1999a,b), certainly requires for their description more than one
parameter. Nevertheless, a good level of approximation may be attained
with a minimum number of parameters when the parameters are
appropriately defined}. From the numerical and seminumerical
simulations we find that the ratio between the halo radius and the
radius containing 1/5 of the total mass,
\begin{equation}
c_{1/5}=\frac{r_h}{r_{1/5}}=\frac{r_h}{r(M_h/5),}
\end{equation}
\noindent is a reasonable
estimator of the halo concentration for most of the haloes. This is
because, for the typical haloes (those with $
\beta \approx 2.7-3.0),$ $r_{1/5}$ is near to the radius $r_m$
where commonly the mass profiles differ more from one to another. The
parameter $c_{1/5}$ correlates with $c_{\rm NFW}$ for a given value of
$\beta $. In Figure 7 we plot these two parameters for halo profiles
from the numerical simulations with $\beta $ $\approx 2.5$ (stars),
$\approx 3.0$ (dots), and $\approx 4.0$ (crosses). This plot
shows the limitation of the parameter
$c_{NFW}$ when the NFW shape is generalized to the profile given by
eq. (2).  Our results show that $c_{1/5}$ and $\beta $  are
weakly correlated. For practical purposes $c_{1/5}$ and $\beta $ may be
considered as two independent parameters; each one is associated with
different characteristics of the mass distribution of the haloes. While
$\beta $ describes the density profiles at the outer regions, $c_{1/5}$
deals with the overall mass distribution down to intermediate ($\approx
0.5-1.0r_m$) radii.

In Figure 8 we plot the
parameter $c_{1/5}$ as a function of virial mass for the isolated haloes (a),
the haloes in groups and galaxies (b), the haloes in cluster (c), and
the haloes obtained in the seminumerical simulations (d) (in this case
$r_h=r_v$ always). The average values of $c_{1/5}$ at each mass bin were 
used in these plots. We find that the dispersions of $c_{1/5}$ 
have an approximate normal distribution. The
standard deviations corresponding to each bin are presented in Figure 8 with  
the dashed lines. The dispersion in the concentration is related to the 
dispersion in \Vc. According to the mass-velocity relation (see $\S 5.2$),
a halo with $\sim 200$ particles (our lower limit) have $\Vc \approx 
125-130\kms $. Nevertheless, due to the dispersion, there are haloes of this 
mass with smaller or larger circular velocities. Thus, in order to avoid 
statistical incompleteness in the estimate of the dispersion of the 
concentration at masses closer to the lower limit, we have fixed the 
lower limit on velocity to 100 \kms~ instead of 130 \kms.  The thin solid 
lines in each panel are the linear regressions to all the haloes of 
the corresponding sample. We find that the low mass haloes tend to 
be more concentrated than the high mass haloes. The haloes in clusters, 
although with a significant dispersion, also tend to be more 
concentrated than isolated haloes. This is in qualitative agreement 
with Bullock et al. (1999). Since $c_{1/5}$ is a parameter completely
independent of the fitting, the fact that haloes in clusters have
larger values of $c_{1/5}$ than the isolated haloes of the same mass,
suggests that the former have steeper density profiles than the
latter. 

Figure 8 shows that the standard deviations of the concentration 
$c_{1/5}$ are of the order of the whole variation of this parameter 
with mass in the galaxy-mass range. This variation should be taken 
into account by analytical and semianalytical works on galaxy 
formation and evolution. 

The $c_{1/5}-M_h$ relation predicted by the seminumerical approach
is in excellent agreement with the results for the isolated
haloes. Nevertheless, we should note that this agreement is expected. As 
was pointed out in $\S 2.2$, in the seminumerical approach we
have to fix a parameter, $e_0$, related to the ellipticity of
the particle orbit. Here, we have fixed $e_0$ to the value for which the density
profile of a model of  $M_h=10^{12}M_{\odot}$ (produced with
the average MAH) agrees with the profile of a typical isolated halo of
the same mass. Note, however, that the trend of $c_{1/5}$ with the mass and 
its scatter predicted with
the seminumerical approach, are independent from the normalization. 
If we fix $e_0$ using the profile calculated with the procedure outlined 
in NFW (1997) then the concentrations $c_{1/5}$ are smaller than those 
obtained with the profiles of our numerical simulation by a factor 
$\approx 1.3$.  Indeed when we fit the density profiles from the numerical 
simulation to the NFW profile (eq. (1)) and define the virial radius in 
the same way NFW did it (the radius where the average density of the halo 
is 200 times the {\it critical} density), we obtain that the average 
values of the parameter $c_{\rm NFW}$ are $\approx 1.2-1.4$ larger than the 
values calculated with the NFW 1997 procedure. Recently, using 
high-resolution $N$-body simulations, Moore et al. (1999)  have also reported 
values of $c_{\rm NFW}$ $50\%$ higher than the values given in NFW97.

\subsection{The mass-velocity relation and its dispersion}

The average $V_m$ corresponding to several mass bins for isolated
haloes (a), cluster haloes (b), and the haloes obtained in the
seminumerical simulations (c), are plotted vs. the mass in Figure
9. The dashed lines represent the respective standard deviations, and
the thin solid lines are the linear regressions for all the haloes of
the corresponding sample. As previous works have shown (e.g., NFW97;
AFH98; Bullock et al. 1999), a strong correlation of the form
$M_h\propto V_m^\alpha $ at galaxy scales is found. The average slope
$\alpha $ we find for isolated and cluster haloes is $\sim 3.29$ and
$\sim 3.50$, respectively. In the case of the seminumerical
simulations (to be compared with the isolated haloes), $\alpha \sim
3.22$. Again, the numerical and seminumerical approaches give a
similar result.  The $M_h-V_m$ relation exhibits a dispersion which is
due to the statistical nature of the primordial density fluctuation
field (AFH98; Avila-Reese 1998).

 In Figure 10 we present the fractional
rms deviations of the velocity $\sigma _V/<\Vc>$ as a function of mass
for the isolated haloes and the haloes in clusters as well as for the
haloes in the seminumerical simulations. The haloes in clusters have
larger deviations than the isolated haloes. Note, however, that due to
the small number of haloes in clusters, the noise in the determination
of their deviations is high. In the clusters the haloes are subject to
tidal stripping which is able to change the original structural
properties of the haloes (Klypin \etal 1999). These changes introduce
an extra scatter in the mass-velocity relation. The deviations
obtained with the seminumerical approach are very similar to those of
the isolated haloes. The results for a $\Lambda $CDM model of
Eisenstein \& Loeb (1996), who used a very simplified method to
estimate the deviations of the $M_h-V_m$ relation, are also very
similar to those obtained here from the numerical simulations. The
fractional rms scatter in velocity may be translated into the
logarithmic rms deviation of the mass: $\Delta
$log$M_h=\alpha$log$(1+\sigma _V/<\Vc>)$, where $\alpha$ is the slope
of the $M_h-V_m$ relation. For $M_h\approx 10^{12}h^{-1}M_{\odot}$ we
obtain $\Delta $log$M_h=$0.12, 0.18, and 0.11 for the isolated and
cluster haloes, and for the haloes from the seminumerical simulations,
respectively.

In Figure 11 we present the correlation among the residuals of the
$M_h-V_m$ and $c_{1/5}-M_h$ relations for the isolated and cluster haloes.
For a given mass, the more concentrated are the haloes the larger are
their $V_m$. The scatter in this correlation is larger for the haloes
in clusters than for the isolated haloes. Some haloes deviate from the
correlation among the residuals; they apparently have too large $V_m$
for their concentrations. In fact most of these haloes are those which
were truncated ($\S 2.1$); their masses and radii are smaller than the
virial mass and radius, while their velocities and radii where is
contained 1/5 of the mass remain nearly the same. As one may
see from Figure 11 the logarithmic deviations from the
$M_h-V_m$ relation are roughly a factor 2 smaller than the
corresponding deviations from the $c_{1/5}-M_h$ relation. 

\section{Discussion}

\subsection{The density profiles of haloes in clusters and groups}

In $\S \S$ 4.2 and 5.1 it was shown that the outer
shape and the concentration of the halo density profiles appear to be
 influenced by the
environment. As in previous studies (Ghigna \etal 1998; Okamoto \& Habe
1998; Klypin et al. 1999), we also found that the haloes in clusters typically 
have steeper outer slopes than the NFW profile. Naively, one could
expect that haloes in groups have outer profile slopes flatter than
those ot the haloes in clusters, but still steeper than the slopes of
the isolated haloes. Our analysis shows that this is not the case. 
The satellite haloes in groups and galaxy-size systems as well as
the pair haloes, have typically flatter outer profile slopes and are
less concentrated than the isolated haloes. Therefore,
the halo density profiles do not follow a continuous trend along the
cluster-group-field  sequence. This result suggests that the
differences between clusters and groups can not be 
viewed as a simple sequence in density.

Why does the outer density structure of the galaxy-size DM haloes depend
on environment? Tidal stripping plays an important role for haloes
inside clusters: haloes that have been subject to tidal stripping have
steeper outer density profiles than the NFW profile, while the haloes
recently accreted onto the cluster have profiles in agreement with the
NFW profile (Ghigna \etal 1998; Okamoto \& Habe 1998; Klypin \etal
1999).  This might be the case for some of our haloes. The MAH also
influences the structure of the halo. We
find that many of the haloes in clusters could have outer density
profiles steeper than $\beta \approx 3$ because they have more
concentrated profiles than the isolated haloes, and this might be 
because they formed earlier than the latter. For example, in the
range of masses of $4\times 10^{11}\hMsun -5\times 10^{11}\hMsun$, the
density in the central bins of haloes in clusters is typically 1.5-2.0 times
larger than in the case of isolated haloes. Since in both cases the mass
is roughly the same, then the external profile slope should be steeper
for the cluster haloes than for the isolated ones. It should be taken
into account, however, that if the haloes in clusters were tidally stripped,
then their original masses have been decreased by the stripping. This
could also explain why the central densities of present-day haloes in
clusters are larger than those of the isolated haloes of the same
mass. It seems that haloes in clusters tend to have steep outer
density profiles due to both effects: (i) because they formed earlier
than the isolated haloes in such a way their density profiles result
more concentrated than the profiles of the isolated haloes, and (ii)
because their outer parts were affected by the tidal stripping. Note that
in the latter case the original halo concentration $c_{1/5}$ should
be smaller. This is because the total mass of the halo $M_h$ decreases
only as $\sim \ln r_h$ (roughly $M(r)\propto \ln r$ at the outer halo
parts). Thus, the halo radius $r_h$ is truncated due to the tidal
stripping while $r_{1/5}$ remains approximately the same.
 
In groups, which are smaller and less dense than clusters, the tidal
stripping is not a significant process, and the typical epochs of
formation of haloes in groups do not differ much from those of the
isolated haloes. Therefore, the structure of the haloes in groups is not
substantially affected neither by tidal stripping nor by the epoch of
formation. Probably, the effects of recent aggregation and interactions 
between the group members are more important than the stripping for group
haloes. For galaxy-size haloes in clusters the roles of the two processes
are reversed (Okamoto \& Habe 1998). The profiles of
some haloes in groups could be shallower than the equilibrium NFW shape
because the halo is caught just when it begins to share the particles
with a nearby companion. It is also possible that, even after a long
time of virialization, due to the merging of substructures, the
particle orbits are more circular than in the case of ``unperturbed''
haloes (isolated). Therefore, these particles do not penetrate to the
central regions and the density profile is shallower than for the
unperturbed haloes. These situations are even more probable for the
galaxy satellite and pair haloes. The correct answer to the question of
how and why do the structures of the DM haloes depend on the
environment have to come from a careful analysis of their evolution in
different environments. This work is currently in progress.

\subsection{The origin of the mass-velocity relation}

The DM haloes exhibit a tight power-law relation $M_v-\Vc^{\alpha}$
between their masses and maximum circular velocities ($\S 5.2$) with
the slope ${\alpha}\approx 3$. It appears that the shape of
the power spectrum of primordial perturbations is responsible for the
slope.  The power spectrum of fluctuations of the CDM models is such
that the concentration of the DM haloes only slightly depends on mass.
Let us analyze the NFW profile (eq. 1), which describes well the density
profiles of a large fraction of haloes in the numerical and
seminumerical simulations.  The maximum circular velocity \Vc~ of a
halo is equal to: 
\begin{equation} 
 \Vc^2=\frac{GM(<r_m)}{r_m},
\end{equation} 
\noindent where 
\begin{equation}
 r_m\approx 2.16\frac{r_v}{c_{\rm NFW}}, 
\end{equation}
\noindent is the radius at the maximum circular velocity $V_m=V(r_m)$. 
Integrating the NFW profile
up to the radius $r_m$ we find:
\begin{equation}
   M(<r_m)\approx 0.467{M_v}/f(c_{\rm NFW}),
\end{equation} 
\noindent where $f(x)\equiv \ln (1+x) - x/(1+x)$, and the
concentration parameter $c_{\rm NFW}=r_v/r_s$ is a weak function of
the virial mass $M_v$ (e.g., NFW). According to the definitions introduced
by NFW, the virial radius $r_v$ of a halo identified at the present epoch is 
related to its virial mass as follows:
\begin{equation} 
r_v\propto M_v^{1/3}. 
\end{equation} 
\noindent From this relation and from
eqs. (4), (5), and (6) we obtain for the $\Lambda$CDM model that:
\begin{equation} 
  \Vc\approx 6.2\times 10^{-3}\left(\frac{M_v}{h^{-1}M_{\odot}}\right)^{1/3}
        \sqrt{\frac{c_{\rm NFW}} {f(c_{\rm NFW})}} \kms
\end{equation} 
\noindent If $c_{\rm NFW}$  would not depend on
mass, one would have $M_v \propto \Vc^3$. From the fittings of our
halo profiles to the NFW profile and using the same definition of
virial radius as in NFW, we find approximately the dependence of 
 $c_{\rm NFW}\propto M_v^{-0.095}$. Substituting this dependence 
in eq. (8) we find that:
\begin{equation}
   M_v\approx 5.2\times10^4\left(\frac{\Vc}{\kms}\right)^{3.2}h^{-1}M_{\odot } 
\end{equation} 
\noindent This relation is in good
agreement with the results obtained in our numerical and seminumerical
simulations (Fig. 9).

An intuitive (although only approximate) explanation for the halo
mass-velocity relation may be given using simple scaling
relations. For example, Gott \& Rees (1975) predicted that for objects
forming instantaneously (monolithic collapse) from density fluctuations 
with a power-law power spectrum the circular velocity and the density of 
the objects should scale as $M^{\frac{1-n}{12}}$ and $M^{\frac{-3-n}{2}}$, 
respectively, where $n$ is the slope of the power spectrum. For the
CDM models at galactic scales $n\approx -(2.0-2.3)$. Thus, according
to this crude analysis the mass scales as $\Vc^{3.6}$ and  $\Vc^{4.0}$,
respectively, which is
steeper than what we find in our simulation.  The circular velocity in
this approximation is the velocity at the virial radius of the halo.
Note that if the halo density does not dependent on the mass, then the
mass scales as the cube of velocity.  In this instantaneous
approximation the epoch $z_c$ at which the object collapses is related
to its mass $M$ as $(1+z_c) \propto M^{-a}$, where $a=(3+n)/6$, $a
=1/6$ for $n=-2$.  However, in the hierarchical formation scenario the
DM haloes do not form instantaneously; they form in a course of
aggregation of subunits and accreting material. Moreover, due to the
random nature of the primordial density fluctuations, the MAHs for a
given present-day halo mass have a dispersion. Nevertheless, one may
still define a typical epoch of formation of haloes of a given
present-day mass (Lacey \& Cole 1993). For example, this epoch can be
defined as the average value of the redshifts at which the haloes of a
given present-day mass $M$ attain half of its mass.  Using the
extended Press-Schechter approximation we calculate this epoch for
several masses in the range of galaxy masses and for the $\Lambda $CDM
model used here. We obtain:

\begin{equation} 1+z_c(M) \propto M^{-a},
\end{equation} 
\noindent with $a\approx 1/22-1/28$, i.e. the slope of
the collapse redshift-mass relation is much flatter than in the case of
the instantaneous collapse (see also AFH98).  This implies that the densities 
of the haloes are also less dependent on mass than in the case of the
instantaneous collapse. Therefore, the slope of the mass-velocity
relation is smaller than in the latter case, and closer to three, which
agrees better with our numerical results.

Assuming the spherical top-hat collapse model and assuming that the
radius of the virialized object is half of the maximum expansion radius,
we find that (e.g., Padmanabhan 1993): 
\begin{equation} V \propto
(1+z_c)^{1/2}M^{1/3}.  
\end{equation}
\noindent Thus, if the mass-velocity
relation is of the form $M\propto V^{\alpha}$, then
\begin{equation} 
  1+z_c \propto M^{2/\alpha-2/3}. 
\end{equation}

Comparing this expression with eq.(10), we find that the value 
for the slope $\alpha$
is $\approx 3.2$. This is roughly the value we obtain in the
numerical and seminumerical simulations for the $\Lambda $CDM model.
In this simplistic analysis we have not considered the structure of the
haloes; only global scaling laws were used. Nevertheless, the analysis
clearly shows that the power-law relation between mass and circular
velocity (defined at the virial radius) of the haloes is explained by
the power spectrum of fluctuations and the extended (hierarchical)
process of formation of the dark haloes.  The latter on average also
depends on the power spectrum, while the scatter in this process (in
the MAHs) is determined by the statistical nature of the density
fluctuation field.

\subsection{Is the Tully-Fisher relation a direct imprint of the $M_h-V_m$
relation?}

We address the question of what constraints can be obtained by
contrasting the observed Tully-Fisher relation (TFR) for galaxies and
the $M-\Vc$ relation for haloes. As it is well known, the luminosity in 
the infrared passbands ($H$ or $I$ for example) is a good tracer of the
stellar disc mass (e.g., Pierce \& Tully 1992). Therefore, in the 
assumption that the disc mass
is proportional to the total halo mass $M_h$, we would expect that 
the infrared-band TFR is an imprint of the $M_h-V_m$ relation. By
comparing the slopes of the $M_h-V_m$ relation for isolated halos
which we obtained in our
numerical and seminumerical simulations ($\S 5.2$), with the observed
TFR slopes (e.g., Gavazzi 1993; Peletier \& Willner
1993; Strauss \& Willick 1995; Willick \etal 1996; Giovanelli \etal
1997), we find that indeed this seems to be the case. In other 
words, the $M_h/L$ ratio in the infrared bands should not
depend on mass (luminosity). Otherwise the slope of the TFR would
become different from the slope of the $M_h-V_m$ relation ($\approx
3.2-3.3$) which already is in good agreement with the observational
data.
 
 Regarding the deviations from the $M_h-V_m$ relation, they will
contribute to the scatter in the TFR. Observational
estimates indicate a scatter in the TFR of about 0.20-0.45 magnitudes
(e.g., Bernstein \etal 1994; Mathewson \& Ford 1994; Willick \etal
1996; Giovanelli \etal 1997). Assuming again a constant $M_h/L$ ratio, 
these estimates correspond to a scatter in the $M_h-\Vc$ relation 
of $\Delta $log$M_h\approx 0.08-0.18$. These values are in agreement 
with those we find in our numerical and seminumerical simulations.  
For example, for $M_h\approx 10^{12}h^{-1}M_{\odot}$ we find 
$\Delta $log$M_h=$0.12 and 0.11 for the isolated haloes in the 
N-body and seminumerical simulations, respectively.
As Mo, Mao, \& White (1998) and AFH98 noted, the scatter in the TFR is
caused not only by the scatter in structure of the DM haloes (due to the
scatter in the MAHs), but also by the dispersion in halo's spin
parameter $\lambda$. Nevertheless, Firmani \& Avila-Reese (1999b) have
shown that the quadratic contribution of this latter to the total scatter 
of the TFR is small -- only about $25\%$ compared to $75\%$ contributed 
to the scatter by differences in the MAHs. 

 Our conclusion is that {\it the slope as well as the scatter of the 
$M_h-V_m$ relation of the CDM halos are similar to those of the
observed TFR and its scatter in the infrared bands}. This coincidence
suggests that the discs formed within the CDM halos have a $M_h/L$
ratio in the infrared bands independent from mass. There is
no room for intermediate astrophysical processes (star formation, 
feedback, gas cooling) able to introduce a dependence of the 
infrared $M_h/L$ ratio with the mass. Models of galaxy formation
and evolution where the fraction of the total mass available for 
forming stars and the star formation efficiency almost do not depend on 
the total mass of the system, are able to predict most of the structural, 
dynamical and luminous properties of disc galaxies, as well as their 
correlations (Firmani \& Avila-Reese 1999a,b,c). The observed 
color-magnitude and color TF relations can be well reproduced by these
models if the luminosity-dependent dust opacity estimated by Wang 
\& Heckman (1996) from a large sample of galaxies is introduced.
Wang \& Heckman have found that the dust opacity
of disc galaxies increaeses with their luminosities. This 
kind of correction also might help to match the predicted 
luminosity function in the CDM models with that inferred from 
observations (e.g., Somerville \& Primack 1998).

As a matter of fact, the evolution of the luminous part of galaxies is 
a very complicated process, which goes beyond the scope of the 
present paper. It is obvious that
only cooling gas can produce stars, and, thus, luminosity is defined by
a complicated interplay between the cooling and heating in the baryonic
component (e.g., White \& Rees 1978; Rees \& Ostriker 1977; Blanton
\etal 1999; Benson \etal 1999). But this does not mean that the amount
and the distribution of the DM are not important. For galaxies
with \Vc =100-300\kms the total luminosity very likely depends on the
total mass of the baryons available for star formation. The latter
correlates with the DM mass. This dependence of the luminosity on the
DM was observed in hydrodynamical simulations which include realistic
cooling, heating, and star-formation processes (e.g., Yepes \etal 1997;
Elizondo \etal 1999; Steinmetz \& Navarro 1999). The correlation exists
because we are dealing with massive haloes of \Vc =100-300\kms for which
the gas cools relatively fast on a dynamical time scale and a large
fraction of the gas is converted into stars.

The situation is different for haloes with smaller mass ($\Vc \le
50\kms)$, which are capable of expelling most of their gas if only few
supernovae are produced, and which can be affected by the intergalactic
ionizing background. It is also different for larger haloes of $\Vc
>300\kms$, which host groups or clusters of galaxies. In this case the
cooling time is long and gas is not converted into stars.  Most of
arguments against a tight $L-\Vc$ relation is for those group- and
cluster- size haloes (e.g., Blanton \etal 1999; Benson \etal 1999). In
this paper we mostly dealt with galaxy-size haloes for which one 
actually may expect that luminosity correlates with \Vc.

\section{Summary and conclusions}

We have analyzed the environmental distribution, the outer density
profiles, and the structural and dynamical correlations of thousands
of galaxy-size DM haloes identified at $z=0$ in a cosmological $N$-body
simulation of a $\Lambda $CDM model.  We have also studied and
analyzed the formation and evolution of DM haloes using an approach based 
on the extended Press-Schechter approximation and on a generalization
of the secondary infall model.  Our main results and
conclusions can be summarized as follows.

{1.} The density profiles of most of the DM haloes in the $N$-body simulation 
(typically resolved only down to radii $0.3-0.8r_m$) are well fitted by
the profile given by eq. (2) with a distribution of the outer slope
$\beta $ such that at the $16\%$, $50\%$ and $84\%$ of the cumulative 
distribution $\beta $ approximately is $2.5$, $2.9$ and $3.9$, respectively. 
The estimated error due to small number of particles is large when 
$\beta $ is large. The slope $\beta$ very weakly anticorrelates
with the mass. Our results confirm that the NFW profile shape describes 
reasonably well the intermediate and outer regions of a large
fraction of DM haloes, particularly the isolated haloes. Our results, however,
show that some fraction of haloes have outer profiles, which
deviate substantially from the NFW shape.

{2.} The distribution of the slope $\beta $ and the halo concentration
$c_{1/5}$ change with the environment in which the haloes are
embedded. In agreement with previous studies, for a given mass we 
find that haloes in clusters
typically have steeper outer density profiles and are more concentrated
than the isolated haloes. Contrary to a naive expectation, we find that
the haloes in galaxy and group systems as well as the haloes with massive
companions, systematically have flatter and less concentrated density
profiles than the isolated haloes. The fact that the halo density
profiles do not follow a continuous trend along the cluster-group-field
sequence suggests that the difference between clusters and groups cannot 
be viewed only as a question of density.

{3.} Approximately $70\%$ of the galaxy-size DM haloes of
$130\kms<V_m<350\kms$ are very isolated systems in the sense that they
are not contained within larger haloes and they do not have massive
companions ($V_{ m}^{\rm comp}>0.7V_m$ or $M_{h}^{\rm comp}>0.3M_h$)
within a radius equal to 3 times their own radii. The $\approx 13\%$ 
of the haloes are contained within larger haloes.  The haloes in pairs or
multiple systems constitute the $\approx 17\%$ of all haloes.

{4.} The parameter $c_{1/5}$ is a good estimator of the halo
concentration, independent of the profile fitting.  The less massive
haloes tend to have larger values of $c_{1/5}$ than the more massive
haloes.

{5.} The galaxy-size haloes exhibit a relation between their
masses and maximum circular velocities, $M_h \propto V_m^{\alpha}$,
with $\alpha \sim 3.3$ and $\alpha \sim 3.5$ for the isolated and
cluster haloes, respectively. This relation may be considered as an 
imprint of the primordial density fluctuation field. For a mass 
of 10$^{12}\hMsun$ the rms fractional velocity deviation 
$\sigma _V/<\Vc>$ from this
relation is $\sim 0.085$ for isolated haloes and $\sim 0.128$ for
cluster haloes. The deviations correspond to $\log\Delta M_h \sim 0.11$
and $\sim 0.18$ for isolated and cluster haloes respectively. The
deviations of the $M_h-V_m$ and $c_{1/5}-M_h$ relations are tightly
correlated. For a given mass the more concentrated haloes have larger
$V_m$.

{6.} The distribution of the parameter $\beta $ obtained with the
seminumerical approach, is similar to the distribution of $\beta $ for
isolated haloes in the $N$-body simulations; the median is at $\beta \sim
2.78$. The $M_h-V_m$ and $c_{1/5}-M_h$ relations and their dispersions
are similar to those of the isolated haloes, too. This agreement
between two completely different methods is encouraging.

To conclude, we have shown that the shapes and concentrations of DM haloes 
exhibit a diversity and systematic dependence on the halo's environment, 
the NFW shape being close to the the average shape and concentration. 
The diversity and dependence on environment can be important in 
shaping the properties of galaxies and their scatter. Therefore, 
studies of galaxy formation and evolution should make an
attempt to account for these effects. 

\section*{Acknowledgments}

V.A. and A.Klypin would like to thank the 
organizers of the Guillermo Haro Workshop 1999 on
"Large Scale Structure and Clusters of Galaxies"
held at the I.N.A.O.E. in Puebla, Mexico
for hospitality while the latest parts
of this paper were being written. Comments
by the referee are gratefully acknowledged.


\newpage

\begin{table}
\caption{Enviromental distribution of DM haloes
with maximum circular velocities $\Vc >130 \kms$.}
\label{tbl-1}
 \begin{center}
 \begin{tabular}{|l|r|r|}  \\ \hline
Environment & Number of haloes & Percentage of haloes  \\ \hline
{Belongs to:\hfill} \\ 
  {\hfill Cluster} & 227 &6.5 \\ 
  {\hfill Group} & 112 & 3.2 \\ 
  {\hfill Galaxy} & 98 & 2.8 \\ \\
{Not in a larger halo:\hfill} \\ 
 {\hfill Isolated} &2456 &70.2 \\ 
 {\hfill Pairs}    &605 &17.3 \\ \\
{Total\hfill} & 3498 & 100 \\ \hline
 
 \end{tabular}
 \end{center}
 \end{table}

\clearpage
\begin{figure} 
\resizebox{\hsize}{!}{\includegraphics{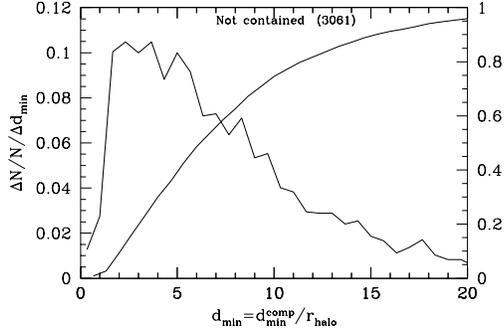}}
\caption{Differential and cumulative distributions of the
 distance from the center of distinct halo (not contained within
a larger halo) to its
nearest significant companion ($V_m^{\rm comp}>0.7\Vc$). The
distance is scaled to the radius $r_h$ of the halo. The
bin width is $\Delta d_{\min}=0.67$.}
\end{figure}

\begin{figure} 
\resizebox{\hsize}{!}{\includegraphics{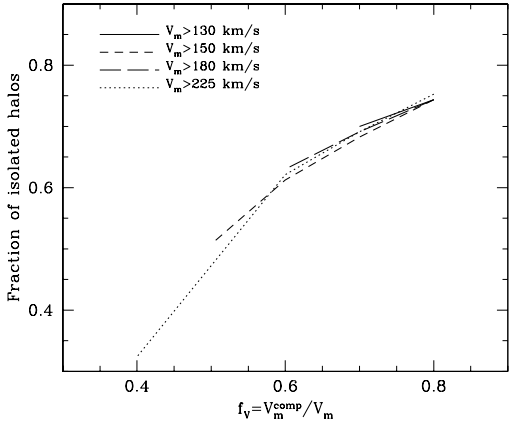}}
\caption{Fraction of isolated haloes in the halo catalog as 
a function of the lower limit 
on the circular velocity of the companion $f_V=V_m^{\rm
comp}/\Vc$. The isolated haloes are haloes not contained within larger
haloes and without any companion with circular velocity
$V_m^{comp}>f_V\Vc$ within 3$r_h$, where \Vc and $r_h$ are the
circular velocity and radius of isolated halo. The lines are for
different samples. The only difference between these samples is the
lower limit on $\Vc$ (shown in the panel) allowed for the isolated
haloes. The curves are truncated at the value of $f_V$ at which the
sample becomes incomplete in the sense that the companions of the
smallest isolated haloes can be smaller than 90\kms.}
\end{figure}

\begin{figure} 
\resizebox{\hsize}{!}{\includegraphics{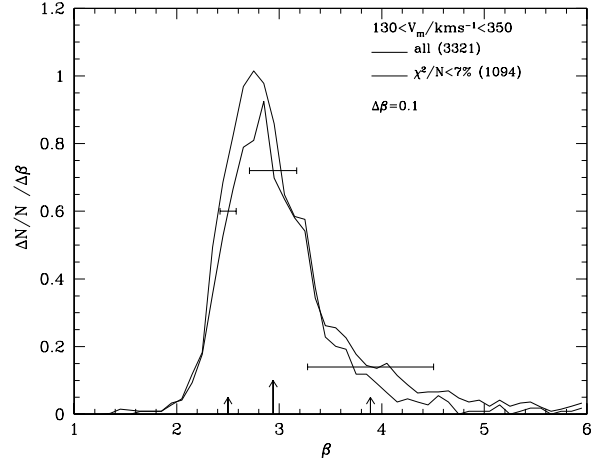}}
\caption{Differential distribution of the slope $\beta $ 
for all the haloes with $130~\kms<\Vc<350~\kms$ (solid line) and only
for the haloes with the low values of $\chi^2$ of the fit (thin solid
line). The arrows indicate the $16\%$, $50\%$, and $84\%$ of the
cumulative distribution of all the haloes. The bars show the errors of
$\beta $.}
\end{figure}

\begin{figure} 
\resizebox{\hsize}{!}{\includegraphics{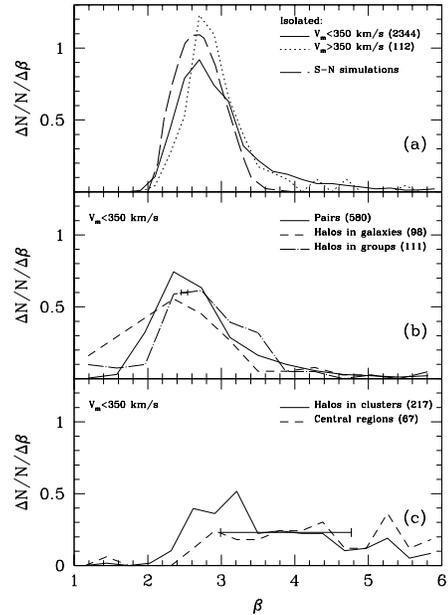}}
\caption{Same as Fig. 3 but for different environments indicated in 
the panels, and for the
haloes obtained in the seminumerical simulations (dashed line in the
upper panel). For the isolated haloes (upper panel) the distribution
for haloes larger than the galaxy sizes, is also shown (dotted
line). The bin widths $\Delta \beta$ used to calculate the
distributions shown in the upper, medium, and lower 
panels were fixed to 0.20, 0.38 and 0.30, respectively. The bars 
plotted for the distribution of the group and cluster haloes show
the error we estimate in the determination of the
parameter $\beta $ when $\beta \approx 2.5$ and 3.9 for the group and
cluster haloes, respectively. }
\end{figure}

\begin{figure} 
\resizebox{\hsize}{!}{\includegraphics{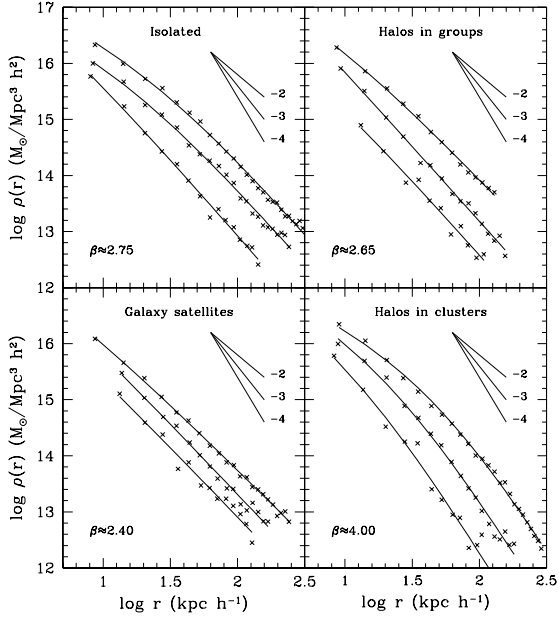}}
\caption{Density profiles of haloes in different
environments (crosses) and the fittings to these profiles using the 
generalized NFW profile given by eq. (2) 
(solid lines). For each sample three haloes were randomly chosen in
three mass ranges 
with $\beta $ around the peak of the distribution of the corresponding 
sample ($\beta $ is shown in each panel). For the haloes in clusters we 
have chosen $\beta =4$ instead of $\beta $ at the peak of the
distribution. The stright lines indicate different slopes.}
\end{figure}

\begin{figure}
\resizebox{\hsize}{!}{\includegraphics{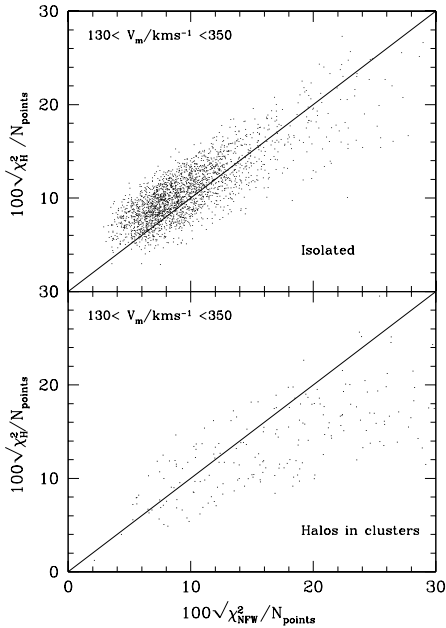}}
\caption{The goodness-of-the-fit parameter 
$\sqrt {\chi^2/N_{\rm bins}}$ in percentages for the cases when the halo
density profiles were fitted to the Hernquist and NFW profiles. The
upper and lower panels are for the isolated haloes and the haloes in
clusters, respectively.}
\end{figure}

\begin{figure}
\resizebox{\hsize}{!}{\includegraphics{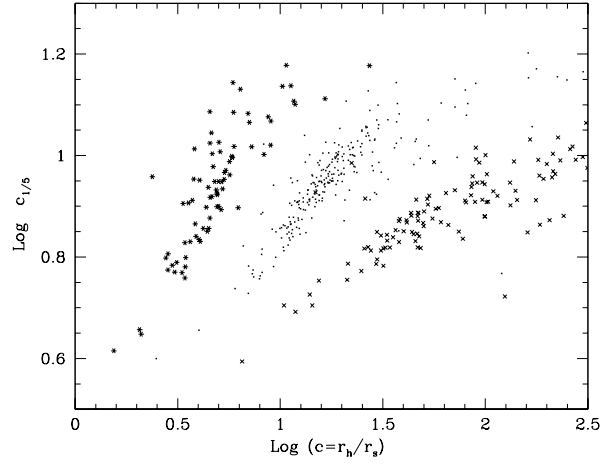}}
\caption{The concentration parameter
$c_{1/5}$ vs. the concentration parameter $c=r_h/r_s$, where
$r_s$ is the scale radius in the fitting formula eq. (2), for
all the haloes with $\beta$ $\approx 2.5$ (stars), $\approx 3.0$
(dots), and $\approx 4.0$ (crosses).}
\end{figure}

\begin{figure}
\resizebox{\hsize}{!}{\includegraphics{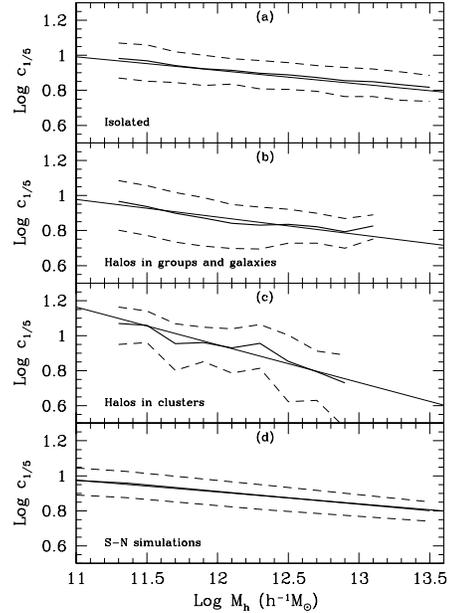}}
\caption{Dependence of concentration $c_{1/5}$ on halo mass
$M_h$ for haloes in different
environments (indicated in the panels), and for the haloes from
the seminumerical simulations (panel d). The solid lines refer to the
average concentration calculated for several mass bins. The
logarithmic widths of the bins are $\Delta M_h=0.10-0.12$. The
standard deviations are represented with the dashed lines. The thin
solid lines are the linear regressions applied to all the haloes of the
given sample.}
\end{figure}

\begin{figure}
\resizebox{\hsize}{!}{\includegraphics{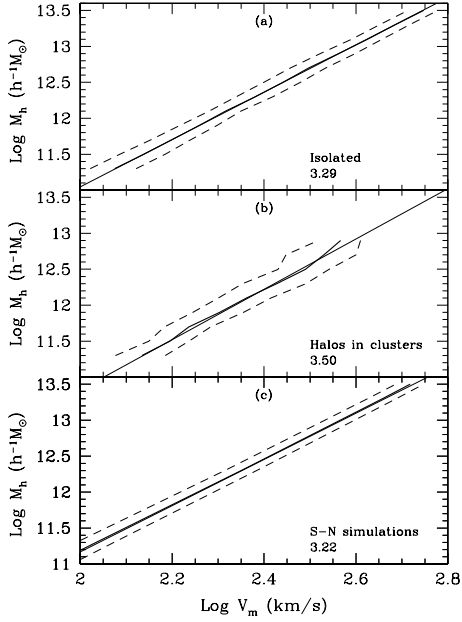}}
\caption{Dependence of mass $M_h$ on the maximum circular
velocity \Vc~ for the isolated haloes (a),
the haloes in clusters (b), and the haloes obtained in the seminumerical 
simulations (c). The same line code of Fig. 8 is used. The 
slopes of the linear regressions (thin solid lines) are indicated
within each panel.}
\end{figure}

\begin{figure} 
\resizebox{\hsize}{!}{\includegraphics{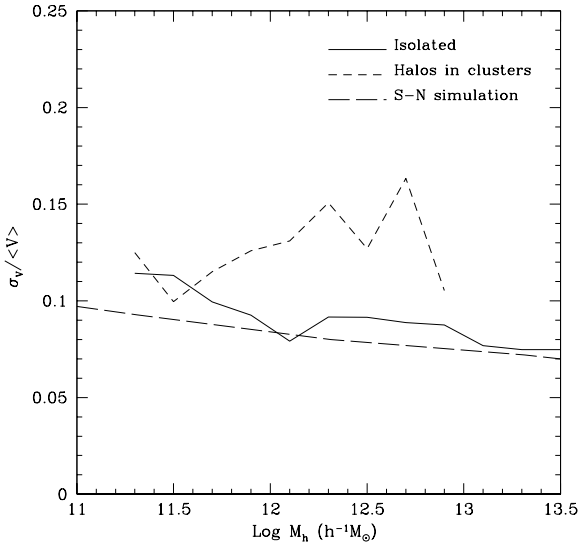}}
\caption{Fractional rms scatter in velocity of the
mass-velocity relation as a function of the mass. The solid and
short-dashed lines are for the isolated and cluster haloes, while the
long-dashed line is for the haloes in the seminumerical simulation.}
\end{figure}

\begin{figure}
\resizebox{\hsize}{!}{\includegraphics{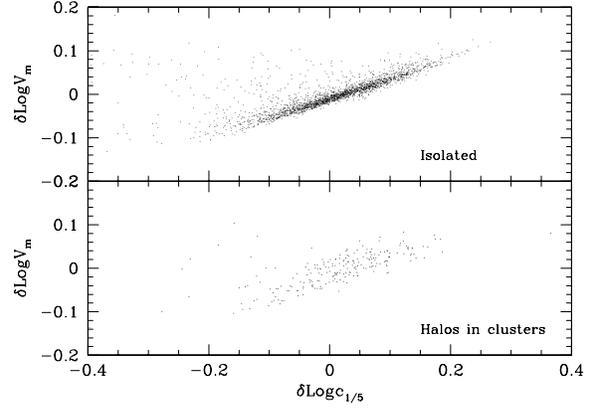}}
\caption{Correlation among the residuals of the $M_h-\Vc$
and $c_{1/5}-M_h$ relations for the isolated haloes (upper panel) and
for the haloes in clusters (lower panel).}
\end{figure}

\end{document}